\documentclass[a4paper,11pt]{article}
\usepackage{aaskaiid}
\usepackage{threeparttable}
\usepackage{graphicx}
\usepackage{soul}
\usepackage{orcidlink}
\usepackage[ragged]{sidecap}
\usepackage{xcolor}

\title{Uncovering neutral Hydrogen clouds in Radio Galaxies in the SKA era}
\ShortTitle{Uncovering H~\textsc{i} clouds in Radio Galaxies in the SKA era}

\author[1]{Mamta Pandey-Pommier\orcidlink{https://orcid.org/0000-0001-5829-1099}}
\ShortName{M. Pandey-Pommier et al.}

\affiliation[1]{Pole Scientific, University Catholic of Lyon- University of Lyon, 10 place des Archives 69288, Lyon, France}
\emailAdd{mamtapommier@gmail.com}

\abstract{
AGN feedback driven by radio jets plays a key role in regulating the cold interstellar medium (ISM) of galaxies. Neutral atomic hydrogen traced through the H~{\sc i} 21-cm line provides a powerful probe of the kinematics, distribution, and physical conditions of cold gas in the central regions of AGN. Previous observations have detected H~{\sc i} column densities down to $\sim10^{20}$ cm$^{-2}$, but typically at arcsecond-scale resolution, inhibiting the characterization of small-scale H~{\sc i} gas clouds and their connection to molecular gas reservoirs and sites of jet--ISM interaction. High-resolution H~{\sc i} imaging is therefore required to determine whether the atomic gas is associated with circumnuclear structures, jet-driven outflows, compressed gas layers, or fragmented cold clouds embedded within a disturbed multiphase ISM. In this chapter, we focus on molecular gas-rich radio AGN hosting large-scale jets and exhibiting strong interactions between the radio plasma and the surrounding ISM, where atomic, molecular, and ionized gas coexist. These systems provide ideal laboratories for investigating the spatial distribution and kinematics of H~{\sc i}, constraining the impact of radio jets on the cold gas, and determining whether the gas is associated with inflowing, outflowing, or otherwise disturbed components. In such environments, H~{\sc i} traces the cold ISM, while extended radio jets and lobes provide structured background continuum emission that enables the atomic gas to be probed across multiple locations within the host galaxy. SKA-VLBI will deliver milliarcsecond-scale imaging and $\mu$Jy-level sensitivity, resolving H~{\sc i} gas clouds on parsec scales across extended radio structures. This capability will enable detailed constraints on the location, morphology, and kinematics of atomic gas from the nuclear regions to kiloparsec-scale jet--ISM interaction sites. Combined with molecular gas observations at other wavelengths, these measurements will provide a comprehensive view of the jet--ISM interactions, the impact of AGN-driven feedback, and the role of cold gas in galaxy evolution.
}

\begin{document}
\maketitle
\section{Introduction} 
Active Galactic Nuclei (AGN) are among the most energetic phenomena in galaxy evolution, where accretion onto supermassive black holes launches relativistic jets that deposit mechanical energy into the surrounding interstellar and circumgalactic medium (ISM and CGM). These jets operate from kiloparsec to megaparsec scales and are a central driver of radio-mode feedback, capable of heating, disturbing, and redistributing the cold gas reservoir of galaxies, thereby regulating star formation and influencing the transition from gas-rich star-forming systems to quiescent galaxies \citep{Silk1998, Fabian2012, Heckman2014, Morganti2017}. The cold ISM in AGN host galaxies is multiphase, consisting of atomic, molecular, and ionized components tracing different physical conditions and evolutionary stages of the gas cycle. Observationally, this multiphase structure is revealed through optical emission-line regions and ionized filaments, infrared dust emission, and molecular gas traced by CO and other dense gas tracers, often showing signatures of disturbed kinematics and outflows \citep{Veilleux2005, Tadhunter2016}. In particular, molecular gas represents the star-forming reservoir, while atomic gas provides a key intermediary phase in the conversion between ionized and molecular components \citep{Krumholz2013, Leroy2008}. These multiphase, filamentary, and clumpy gas structures reflect the dynamic and turbulent conditions within the ISM affected by AGN activity \citep{Veilleux2005, Morganti2017}. 

\begin{figure}[ht]
    \centering
	\includegraphics[width=0.5\columnwidth]{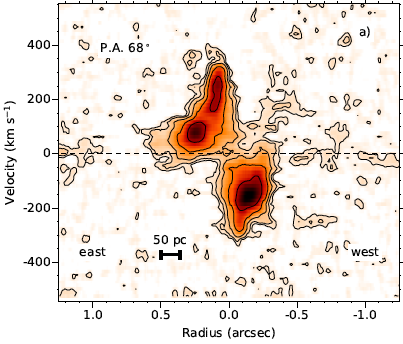}
    \includegraphics[width=0.45\columnwidth]{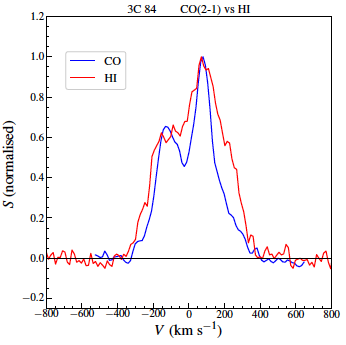}
     \caption{\textbf{Panel 1 \& 2:}Cold gas as observed with ALMA \citet{Oosterloo2023} and H~{\sc i} and CO profiles  measured in 3C84 \citet{Morganti2023}.
}
\label{fig:3C48-multigas}
\end{figure}

Neutral atomic hydrogen H~{\sc i}, traced through the 21-cm transition, is particularly valuable because it probes cold gas kinematics in the vicinity of the AGN, providing insight into the circumnuclear regions, where inflow-outflow processes take place. H~{\sc i} is commonly detected in both emission and absorption in radio-loud AGN, revealing rotating disks, inflowing material, and jet-driven outflows of a few kiloparsec or smaller in size \citep{Morganti2005, Oosterloo2025}. Large-area surveys with SKA pathfinders such as MeerKAT, ASKAP, GMRT, and the VLA have significantly expanded the census of H~{\sc i} in radio AGN at low redshift ($z \lesssim 0.25$), revealing complex line profiles and disturbed gas kinematics, see Fig.~\ref{fig:3C48-multigas} \citep{Maccagni2023, Morganti2017, Vermeulen2003}. Associated H~{\sc i} absorption is detected in $\sim$28\% of nearby radio AGN, with higher incidence and broader, more complex profiles in compact sources \citep{Maccagni2017}. These observations suggest that the distribution and kinematics of the absorbing H~{\sc i} gas may depend on radio power, the radio continuum emission, and the dust content of these sources. These findings are consistent with turbulent, jet-disturbed environments containing substantial reservoirs of cold gas that impact both AGN fuelling and feedback. However, most existing observations remain spatially unresolved or only marginally resolved, making it difficult to distinguish between circumnuclear structures, outflowing gas, fragmented clouds, and material displaced by jet--ISM interactions. Consequently, the spatial distribution of H~{\sc i} gas relative to radio jets, sites of jet--ISM interaction, and molecular gas reservoirs remains poorly constrained, particularly in molecular-gas-rich radio AGN where multiple gas phases coexist.

High-resolution very long baseline interferometry (VLBI) studies have demonstrated that spatially resolving H~{\sc i} absorption can directly identify jet–cloud interactions on parsec scales, as shown in systems such as IC~5063, 4C~31.04, and 3C~236 \citep{Morganti2003, Morganti2004, Schulz2018}. These studies reveal that radio jets can strongly disturb cold gas, producing localized outflows, shocks, and kinematically complex absorption structures. However, such observations remain limited to small samples due to sensitivity and frequency constraints. Further, the interactions encountered by the jet within the ISM are often imprinted in its morphology as bent or distorted asymmetries with wiggles, knots, or disrupted lobes that can be traced via continuum observations. These structural features are sites of strong interactions with the ambient medium, leading to shock heating, gas compression, and potentially localized star formation regions \citep{Laing2014}. In contrast, in lower-density environments, jets tend to expand symmetrically and traverse greater distances with minimal disruption. Thus, combined H~\textsc{i} and radio continuum studies are essential for understanding the coupling between AGN jets and the gaseous ISM, the evolution of their host galaxies, and the regulation of star formation \citep{Wiklind1995, Cicone2014, Pandey-Pommier01.2026.SKA}. Furthermore, at higher redshifts ($z > 0.6$), H~{\sc i}  detection rates decrease, primarily due to observational selection effects, including limited sensitivity, luminosity, and source morphology biases, as discussed in the comprehensive review on HI absorption surveys by \citep{Curran2024}. Further, recent pathfinder surveys such as ASKAP-FLASH and MeerKAT-MALS have significantly reduced selection biases compared to earlier targeted studies, providing the first statistically meaningful constraints on the evolution of cold gas in radio AGN \citep{Aditya2024, Gupta2017}. However, sensitivity to weak or broad absorption features and dependence on background continuum sources still impose residual selection effects. Early targeted H~{\sc i} absorption studies were biased toward flat-spectrum radio sources due to the selection of compact, core-dominated targets \citep{Aditya2018, Aditya2019}. However, recent pathfinder surveys such as ASKAP-FLASH and MALS have significantly reduced this bias by adopting more uniform or blind selection strategies; however, high-redshift detections remain rare, and unresolved absorption profiles continue to limit the interpretation of gas kinematics and geometry. In addition, the connection between the atomic and molecular gas phases and sites of jet--ISM interaction remains poorly constrained. Although VLBI follow-up observations have demonstrated the power of spatially resolving H~{\sc i} absorption in individual systems, revealing compact clouds and localized jet--cloud interactions in a small number of nearby radio galaxies, such studies remain limited by sensitivity and the restricted frequency coverage currently available.

In this chapter, we investigate molecular gas-rich radio AGN hosting large-scale FR~I/II jets with radio powers as high as $P_{1.4,\text{GHz}} \sim 10^{26},\text{W Hz}^{-1}$ as a key population for studying jet-ISM interactions in multiphase environments. A particularly relevant subset is provided by molecular hydrogen emission galaxies (MOHEGs), which are well-known examples where strong radio jets coexist with abundant molecular and atomic gas, yet star formation remains significantly suppressed \citep{Ogle2010}. In these systems, H~{\sc i} absorption is detected in a large fraction of sources ($\sim 70\%$), but shows no clear correlation with molecular gas mass or star formation rate, suggesting that the presence of cold gas alone is not sufficient to sustain star formation \citep{Ogle2010, Wagh2024}. This points to strong feedback effects that regulate or disrupt the cold gas on small scales, potentially inhibiting star formation despite substantial gas reservoirs. Such processes can be directly investigated using VLBI H~{\sc i} 21-cm observations, which resolve the absorption structure on parsec scales and can identify localized jet-cloud interactions, outflows, and shocked gas within the ISM, responsible for star-formation quenching in these systems.

To localize outflows and trace their structure, H~{\sc i} mapping at sub-arcsecond angular resolution is required. Previous studies have detected H~{\sc i} absorption with column densities as low as $10^{20}~\mathrm{cm}^{-2}$ in AGNs; however, the combination of mJy-level sensitivity and angular resolutions of only a few tens of milliarcseconds has limited our ability to probe the small-scale structure of the H~{\sc i} gas and to determine whether it is associated with circumnuclear regions, jet–ISM interaction sites, or unrelated foreground material along the line of sight. In particular, the spatial distribution of atomic gas-whether confined to circumnuclear structures, fragmented into clouds embedded in a turbulent medium, or compressed and accelerated by jet-driven shocks-remains poorly constrained. Addressing these questions requires spatially resolved H~{\sc i} observations that can directly link the distribution and kinematics of the atomic gas to both the radio continuum morphology and the molecular gas reservoirs as seen in 3C326 EVN observations \citep{Schulz2018}. The SKA-VLBI will provide this capability by combining milliarcsecond angular resolution with $\mu$Jy-level sensitivity. This unprecedented combination will enable spatially resolved measurements of H~{\sc i} gas clouds, allowing the spatial distribution and kinematics of the atomic gas to be traced along multiple sightlines through the host galaxy. By comparing the spatial correspondence between atomic gas clouds and the resolved radio continuum, SKA-VLBI observations will help differentiate between likely scenarios, including (i) gas tracing circumnuclear regions (including rotating disks and potential inflow components), (ii) compressed gas aligned with jet or lobe structures indicating jet–ISM interaction regions, and (iii) spatially offset and fragmented gas clouds or unrelated foreground material. When combined with spatially resolved molecular gas observations (e.g., CO from ALMA), these data will allow direct tests of whether atomic gas is co-located with molecular reservoirs or instead occupies shocked, ionized-to-molecular transition layers produced by jet–ISM interactions.  Furthermore, SKA-VLBI will enable to investigate the spatial association of H~{\sc i} gas clouds with radio jets, molecular gas reservoirs, and feedback-affected regions, thereby providing constraints on the physical conditions governing the atomic-to-molecular gas transition, the origin and fate of inflowing and outflowing gas, and the role of cold gas in regulating galaxy evolution. The central goal is to determine how AGN-driven radio jets impact the redistribution, acceleration, and multiphase-transformation on parsec-to-kiloparsec scales.

\section{VLBI Studies of H~\textsc{i} Outflows in AGN}
\begin{figure}[h]
    \centering
	\includegraphics[width=0.275\columnwidth]{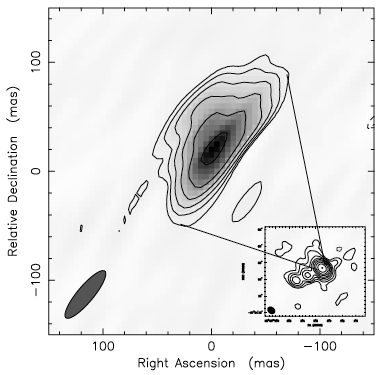}
    \includegraphics[width=0.39\columnwidth]{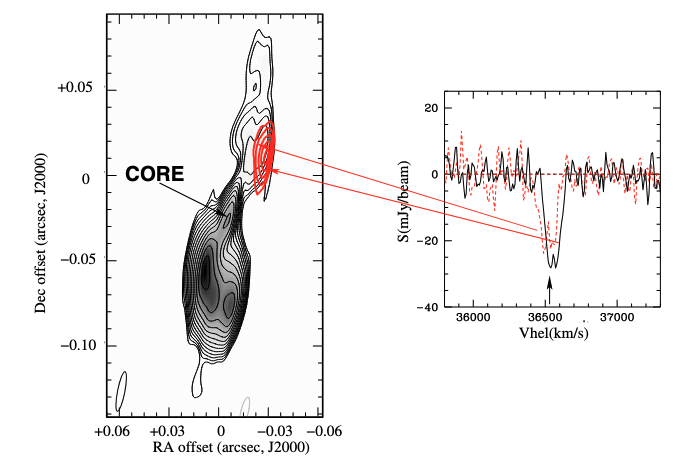}
    \includegraphics[width=0.32\columnwidth]{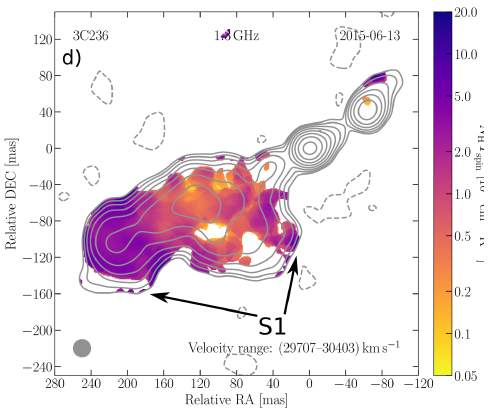}
     \caption{\textbf{Panel 1 \& 2:}HI gas outflows as observed with VLBI in IC 15063 \citep{Oosterloo2023} and 4C 12.50 \citep{Morganti2004}. \textbf{Panel 3:} 3C326: Map of N(H~{\sc i})$T^{-1}$ spin for H~{\sc i} features detected with the EVN \citep{Schulz2018}.}
\label{fig:VLBI-imaging}
\end{figure}

A big caveat of H~\textsc{i} 21-cm absorption studies is that the observed line properties depend not only on the intrinsic gas distribution but also on the geometry of the background continuum source. Orientation effects, covering-factor variations, and differences between compact and extended radio morphologies can significantly influence both detectability and inferred gas kinematics, introducing important selection biases in low-resolution absorption surveys. These biases are best addressed statistically through large, heterogeneous samples spanning a wide range of radio morphologies and jet orientations. In contrast, VLBI H~\textsc{i} observations are typically conducted toward sources already detected in low-resolution surveys and are designed to spatially resolve the absorbing gas on parsec scales. By separating distinct kinematic components and associating them with specific jet, core, or lobe structures, VLBI provides a direct view of the gas distribution, kinematics, and the physical mechanisms driving jet–ISM interactions. The first clear detection of a spatially resolved H~\textsc{i} outflow was obtained in the nearby Seyfert galaxy IC~5063 by \citep{Oosterloo2023}. VLBI imaging revealed fast blueshifted ($\sim700$~km~s$^{-1}$)  H~\textsc{i} absorption spatially coincident with the western radio lobe at a projected distance of $\sim0.6$~kpc from the nucleus. The morphology and kinematics indicate a strong interaction between the radio jet and a dense molecular cloud, producing a shocked and accelerated atomic gas component distributed around the expanding radio lobe. This provided the first direct evidence that jet–ISM interactions can accelerate atomic hydrogen on sub-kiloparsec scales, even in relatively low-power radio galaxies ($P_{1.4\,\mathrm{GHz}} \sim 6\times10^{23}$ W Hz$^{-1}$, see Fig.~\ref{fig:VLBI-imaging}- Panel~1).

Subsequent studies confirmed that such jet-driven atomic gas structures are common in compact and restarting radio sources. In compact radio galaxy, 4C~12.50, \citet{Morganti2004} used VLBI to resolve a broad ($\sim2000$~km~s$^{-1}$) blueshifted absorption component into spatially distinct structures associated with the southern radio lobe, extending over several hundred parsecs (see Fig.~\ref{fig:VLBI-imaging}- Panel~2). A separate systemic absorption component was found closer to the nucleus, indicating multiple kinematic phases coexisting within the circumnuclear medium. The inferred mass of the outflowing atomic gas is of order $10^5$~M$_\odot$, with mass outflow rates of $\sim10$–$30$~M$_\odot$~yr$^{-1}$ depending on the assumed spin temperature, assuming reasonable spin temperature values \citep{Morganti2005}.

The morphology and kinematics indicated that the outflow was driven by the young, restarting radio jet, which was interacting with a dense circumnuclear medium, but not necessarily connected, and instead could be a tracer of the large-scale medium that surrounds the active nucleus. This study provided a clear VLBI-scale confirmation of jet-driven neutral hydrogen outflows in powerful AGNs and emphasized the importance of spatially resolved H~\textsc{i} gas clouds to distinguish between inflow, disk rotation, and feedback processes. 

A detailed study of the restarted FR II radio galaxy 3C~236 was presented by \citet{Schulz2018}, who combined VLA and global VLBI observations to map the neutral atomic hydrogen outflow with high spatial and spectral resolution. The VLA spectrum revealed a broad, blueshifted absorption wing extending up to $\sim$1000~km~s$^{-1}$ from systemic, consistent with a fast neutral outflow see Fig.~\ref{fig:VLBI-imaging}- Panel~3). The VLBI imaging resolved this outflow into four discrete H~\textsc{i} clouds with radial velocities ranging from $150$ to $640$~km~s$^{-1}$ relative to the rotating disk component. Three of these clouds were projected within $\lesssim$40~pc of the radio core, while the fourth was located near the southeastern jet lobe, at a distance of $\sim$270~pc. The individual cloud masses were estimated to lie in the range $(0.3$–$1.5)\times10^4$~M$_\odot$, assuming a spin temperature of 100~K. The kinematics and proximity to the radio structure strongly suggest that these clouds are part of a jet-driven outflow. Interestingly, only a fraction of the total absorption detected by the VLA was recovered in the VLBI data, implying that some of the outflowing gas may be in a more diffuse or extended phase, undetectable at VLBI scales. This study highlighted the multi-phase, clumpy nature of cold gas in molecular gas-rich AGN hosting large-scale jets and demonstrated the ability of VLBI to pinpoint the launching sites and dynamics of jet-accelerated H~\textsc{i} gas clouds.

Further, VLBI observations in Perseus A, show that a complex integrated H~\textsc{i} absorption lines with a blue-shifted wing does not trace an outflow~\citep{Morganti2023}, while in MrK 231 arcsecond resolution JVLA observations reveal that the outflow is mislocated with respect to the expansion of the jets~\citep{Morganti2016}, and is thus the wind of the AGN responsible for the acceleration and heating of the circum-nuclear gas.  More recent VLBI surveys have extended such studies to additional sources, including 3C~293, 4C~52.37, and 3C~49 \citep{Schulz2018}, 
These observations show compact clouds of cold atomic gas with masses ranging from 10$^4$ to 10$^5$ M$\odot$, often found within the inner 100–300 pc. However, a significant portion of the absorption detected at lower resolution is often resolved out, indicating that outflows also contain diffuse, extended components that current VLBI sensitivity cannot fully capture. In 3C~293, for instance, only a subset of the broad blueshifted wing was detected in VLBI, corresponding to a local outflow rate of $\sim$0.36 M$\odot$ yr$^{-1}$, compared to the total outflow rate of up to $\sim$50 M$_\odot$ yr$^{-1}$ estimated from VLA data \citep{Schulz2018}. Lastly, observations of NGC~1275 (3C~84) reveal H~\textsc{i} absorption features with large velocity offsets (~8120 km s$^{-1}$) along a compact line of sight ($\sim$16 pc) \citep{DeYoung1973,Ekers1976}. These extreme velocity offsets are unlikely to be associated with AGN-driven outflows and instead may trace intervening or foreground absorbing gas, illustrating how VLBI imaging can isolate narrow sightlines and disentangle spatially distinct absorption components. More broadly, these cases highlight the challenges in interpreting H~{\sc i} absorption profiles, which can arise from a combination of inflowing, outflowing, and unrelated fragmented gas components.

\begin{figure}[h]
    \includegraphics[clip,width=1\textwidth]{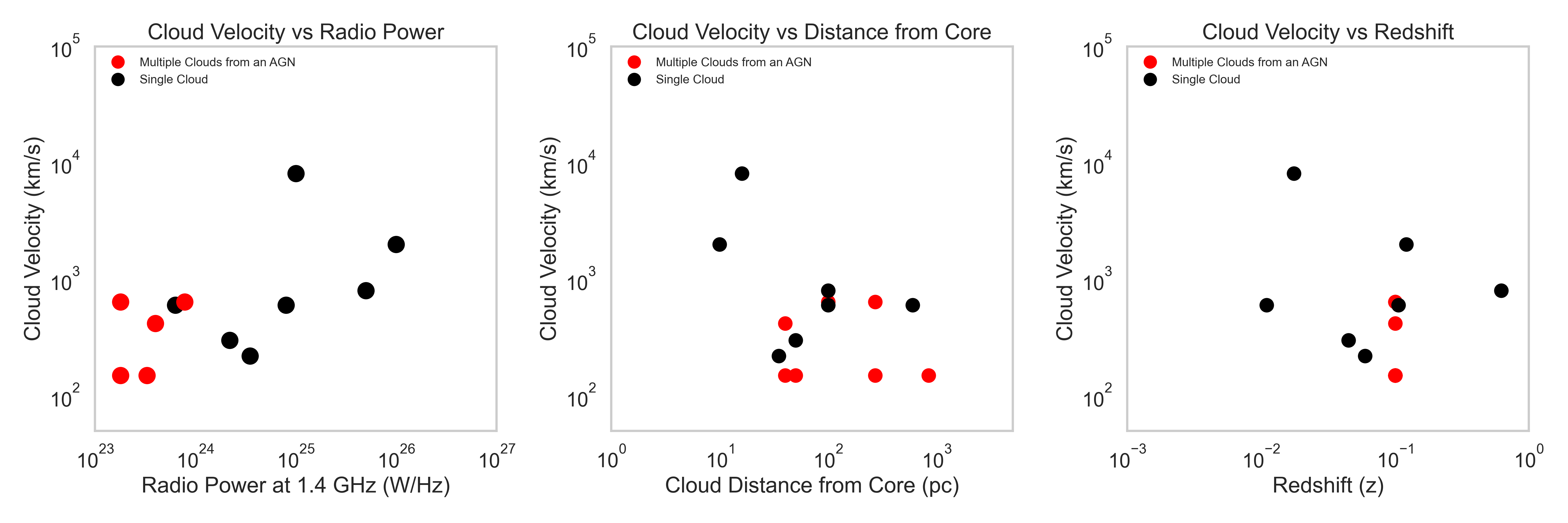}
    \caption{H~\textsc{i} gas clouds and outflows measured via VLBI imaging observations on AGNs.}
\end{figure}

In \textbf{Fig. 3}, we present plots summarizing key properties of H~\textsc{i} outflows imaged by VLBI in various AGN systems, focusing on gas cloud sizes and velocity offsets across different redshifts and radio powers. These plots offer valuable insights into the trends and variations in AGN feedback processes, as well as the influence of AGN jets on the surrounding ISM. \textbf{Panel 1} reveals a broad range of values for both cloud velocity and radio power, reflecting the complex interplay between several factors, including jet morphology, jet power, and their interactions with the clumpy and heterogeneous ISM. While a slight trend suggests higher cloud velocities correlate with higher radio power, the data also show that cloud velocity is not directly tied to the total energy output of the galaxy in the radio spectrum. The wide distribution of data points suggests that the physical processes driving galaxy activity are complex, with both low and high-power radio galaxies exhibiting a broad spectrum of cloud velocities. \textbf{Panel 2} illustrates that galaxies with both high and low cloud velocities can host clouds at varying distances from the galaxy's core, implying that galaxy outflows or jet-driven activity are not purely radial in nature. The cloud movement may be influenced by multiple factors, such as jet orientation and gas dynamics, rather than a simple radial expansion. Notably, gas clouds at higher velocities tend to be located closer to the core, whereas galaxies like 3C 236, which exhibit multiple gas clouds, show a linear spread of clouds at larger distances from the core, indicating a smoother ISM distribution. Finally, \textbf{Panel 3} shows that there is no clear trend between redshift and cloud velocity measured until now. These results clearly suggest that the limited number of VLBI H~{\sc i} absorption studies currently available, all of which are restricted to low redshifts, does not yet permit any assessment of possible redshift evolution in the incidence or properties of high-velocity H~{\sc i} clouds  ($z \ge 0.1$). This limitation is primarily due to the restricted frequency coverage of current VLBI facilities, which prevents systematic extension of such studies to higher redshifts, rather than sensitivity constraints alone. Consequently, most VLBI-based constraints on H~{\sc i} absorption arise from a small, nearby sample, and caution is required when generalising these results to the broader AGN population.

While high-resolution VLBI imaging observations have proven essential for resolving the small-scale structure and location of H~\textsc{i} absorption in radio-loud AGNs, offering direct insights into the mechanisms of AGN feedback and the interactions between radio jets and the surrounding ISM. These studies show that H~\textsc{i} outflows are typically clumpy, compact, and confined to scales of a few hundred parsecs from the central engine, often co-spatial with regions where radio jets interact with dense gas clouds. Collectively, these results highlight the importance of VLBI in probing the kinematics and morphology of cold gas in AGN host galaxies. However, current studies remain limited to a small number of nearby systems ($z \lesssim 0.3$), primarily due to the restricted sensitivity and frequency coverage of existing VLBI facilities. The SKA-VLBI will overcome these limitations by providing the sensitivity and angular resolution required to detect low-column-density H~\textsc{i} absorption across a much broader range of radio powers ($10^{22}$–$10^{26}$ W Hz$^{-1}$). This will enable significantly larger and more statistically meaningful samples of AGNs to be studied. As a result, SKA-VLBI will allow systematic investigations of jet-driven feedback, the spatial distribution of atomic gas, and its connection to molecular gas reservoirs, placing the currently sparse VLBI results into a unified framework for cold gas regulation in AGN host galaxies.

\section{The Role of SKA in Advancing H~\textsc{i} Absorption Studies}
\label{ska_contributions}
\begin{figure}[h]
    \centering
	\includegraphics[width=0.38\columnwidth]{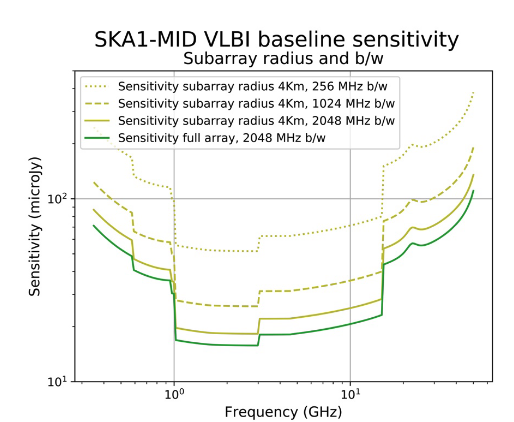}
    \includegraphics[width=0.61\columnwidth]{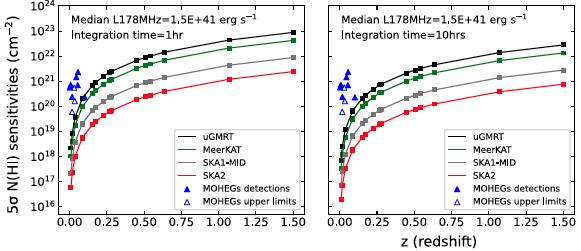}
     \caption{\textbf{Panel 1:}Baseline sensitivity for different SKA1-MID subarrays configurations and a 100m class remote
telescope, for different observing bandwidths - adapted from - \href{https://www.skao.int/sites/default/files/documents/JJ-WP10deliverable10.3.pdf}{https://www.skao.int/sites/default/files/documents/JJ-WP10deliverable10.3.pdf}. \textbf{Panel 2 and 3:} N(H~{\sc i}) sensitivities for SKA and pathfinders w.r.t to detect cold gas in MOHEGs for different integration times \citep{Wagh2024}.}
    \label{fig: VLBI-sensitivity}
\end{figure}

The SKA VLBI will transform our understanding of the cold ISM in galaxies, particularly in detecting and resolving H~\textsc{i} clouds with unprecedented precision. The combination of SKA-Mid's sensitivity, wide bandwidth, resolution, and VLBI capabilities will provide valuable insights into AGN feeding and feedback mechanisms across cosmic time. Specifically, SKA1-MID Bands 1 and 2 (0.35–1.76 GHz) will be capable of probing redshifted H~\textsc{i} absorption from the local Universe to a redshift of $z \sim 3.5$, depending on background continuum brightness and spectral-line detectability (see Fig.~\ref{fig: VLBI-sensitivity}).

\begin{table}[h]
\centering
\begin{threeparttable}

\caption{Expected physical scales of H~{\sc i} clouds for different SKA-VLBI configurations at 1.4 GHz.}
\label{tab:vlbi_hi_scales}

\begin{tabular}{lcccp{6cm}}
\hline
Configuration & Maximum & Scale & Scale & Typical H~{\sc i} \\
 & Baseline & at $z=0.01$ & at $z=0.1$ & Science \\
\hline

SKA-MID Core
& 20 km 
& $\sim$450 pc 
& $\sim$4 kpc 
& Diffuse H~{\sc i} gas, galaxy-scale outflows, extended jet--ISM interaction regions \\

SKA-MID Full Array 
& 150 km 
& $\sim$60 pc 
& $\sim$530 pc 
& Circumnuclear H~{\sc i} structures, compact outflows \\

SKA + African VLBI  
& $\sim$1000 km 
& $\sim$9 pc 
& $\sim$80 pc 
& Localization of H~{\sc i} absorption against compact jets \\

SKA + EVN / VLBA 
& 3000--5000 km 
& $\sim$2--3 pc 
& $\sim$16--27 pc 
& Parsec-scale H~{\sc i} clouds and jet--cloud interaction sites \\

Global SKA-VLBI$^{*}$  
& 8000--10000 km 
& $\sim$1--1.5 pc 
& $\sim$8--12 pc 
& Detailed kinematics of compact H~{\sc i} components in AGN cores \\
\hline
\end{tabular}

\begin{tablenotes}
\footnotesize
\item[${*}$] Global SKA-VLBI includes EVN, uGMRT, VLBA, and LBA. Note that this table is intended as a representative summary; H~{\sc i} detectability with VLBI is ultimately limited by the availability of compact background continuum sources, the accessible frequency coverage, and surface-brightness sensitivity constraints. These factors limit the detectability of diffuse or low optical depth absorption even when angular resolution is sufficient.
\end{tablenotes}

\end{threeparttable}
\end{table}

SKA1-MID will achieve sub-$\mu$Jy sensitivity in deep integrations, enabling the detection of associated H~\textsc{i} absorption down to column densities of $N_{\mathrm{H\,I}} \sim 10^{18}$–$10^{19}\,\mathrm{cm}^{-2}$ for bright background continuum sources, depending on spin temperature and covering factor assumptions. In practice, such sensitivity will enable blind and targeted absorption surveys over large sky areas, significantly increasing the known population of cold gas-rich AGN. The full SKA-MID array, with an angular resolution of $\sim0.3$$"$ at 1.4 GHz, wide frequency coverage, and high sensitivity, will be ideally suited for mapping H~\textsc{i} in nearby ($z \le 0.01$) molecular gas-rich radio AGN hosting large-scale jets. At these distances, the achieved resolution corresponds to spatial scales of a few $10s$ of parsec, enabling the study of the distribution and kinematics of cold atomic gas in circumnuclear regions and sites of jet--ISM interaction. These observations will provide detailed imaging of H~\textsc{i} structures. The high surface-brightness sensitivity provided by the inner 20 km SKA-MID configurations (e.g., AA* and AA4), corresponding to an angular resolution of $\sim0.2$$"$
at 1.4 GHz, will be particularly effective for detecting low-column-density H~\textsc{i} structures on scales of several hundred parsecs in nearby galaxies. This capability will allow diffuse atomic gas associated with outflows, filaments, and extended jets-ISM interaction regions to be traced with unprecedented sensitivity. In contrast, SKA-VLBI observations, in combination with facilities such as the EVN, VLBA, uGMRT, and LBA will achieve resolutions of a few to tens of milliarcseconds, corresponding to parsec-scale imaging in nearby AGN and tens of parsecs at intermediate redshift \citep{Li2024}. This combination of high sensitivity and milliarc second resolution is essential for localizing H~\textsc{i} absorption relative to radio jets and identifying the sites of jet–ISM interaction (see Table 1). This resolution is crucial for the core regions of AGNs, where processes like jet-driven outflows occur, and contain small-scale H~\textsc{i} structures that are vital to understanding galaxy evolution. By mapping H~\textsc{i} clouds in these regions, we can investigate how AGN feedback affects the surrounding gas, how outflows interact with the ISM, and the extent to which AGN activity alters the gas distribution. This will also provide critical insights into the presence of small-scale features such as filaments, clumps, and outflows, which are driven by AGN feedback. When combined with continuum data, we will be able to explore whether jets heat or disperse H~\textsc{i} gas, and how this process regulates star formation in the host galaxy.

\section{Conclusions}
AGN feedback plays a critical role in shaping the evolution of galaxies and their environments. This feedback, particularly via jet-ISM interactions, has significant implications for the dynamics of gas in AGN host galaxies, regulating their star formation efficiencies. H~\textsc{i} outflows, as traced through mapping observations, provide valuable insights into the kinematics and physical conditions of the cold atomic gas in these regions. The role of H~\textsc {i} in AGN feedback, however, remains complex and still poorly understood, particularly in the way it interacts with molecular gas and regulates star formation processes. Recent high-resolution VLBI studies have begun to resolve the small-scale structure of H~\textsc{i} absorption in radio-loud AGN, revealing compact, clumpy gas components and direct evidence of jet–driven perturbations in systems such as IC 5063, 4C 12.50, and 3C 236. These results demonstrate that the observed H~\textsc{i} absorption is strongly dependent on both jet morphology and orientation, highlighting the importance of spatially resolved observations to disentangle rotating disks, inflows, and feedback-driven outflows. However, current VLBI studies remain limited to a small number of nearby systems ($z \lesssim 0.25$), largely due to frequency coverage constraints.

The SKA-VLBI facility promises to revolutionize our understanding of AGN feedback by providing unparalleled sensitivity, frequency coverage,  spatial resolution, and survey speed. SKA’s VLBI capabilities, particularly with its ability to map H~\textsc{i} gas with sub-arcsecond resolution at $\mu$Jy-level sensitivities, will enable detailed studies of gas dynamics in AGNs across a broad range of redshifts. The ability to detect H~\textsc{i} outflows in galaxies up to redshift $z \sim 3.5$, along with its high dynamic range and sensitivity, will provide new insights into how AGNs regulate the surrounding ISM. Through these observations, we will better understand the role of AGN-driven gas outflows in galaxy evolution and their broader impact on cosmic structures. In particular, SKA's ability to probe small-scale H~\textsc{i} clouds features and trace jet-induced interactions will allow us to study the feedback processes that govern galaxy transformation, helping to connect the small-scale dynamics of AGN feedback with the large-scale evolution of galaxies. Ultimately, the SKA will provide a comprehensive framework for studying the impact of AGNs on galactic transformation and evolution, as well as the role of cold gas in these processes, in the era of high-precision observational astronomy. It will offer powerful scientific synergy with upcoming facilities, such as ALMA2040, the WST, the ELT, JWST, and next-generation ground-based observatories, enabling a multi-wavelength, multi-messenger approach to deepen our understanding of these complex phenomena. By combining these multiwavelength data, we will gain a comprehensive picture of the processes that shape the central regions of interacting galaxies, providing a crucial step forward in understanding galaxy evolution and the role of AGN feedback in regulating the interstellar medium.

\section*{Acknowledgements}
We thank the referee for a constructive and timely review of the manuscript. The comments and suggestions helped us improve the focus of this chapter.

\bibliographystyle{abbrvnat-maxbibnames4}
\bibliography{chapter}

@incollection{Pandey-Pommier01.2026.SKA,
author = {Mamta Pandey-Pommier and others},
title = {Tracing gas outflows in molecular hydrogen emission-rich galaxies with the ska},
year = {2026},
publisher = {},
note = {arXiv search: Report number AASKAII/Pandey-Pommier01},
booktitle = {Advancing Astrophysics with the SKA -- II (AASKAII)}}

@article{DeYoung1973,
  author  = {De Young, D. S. and Roberts, M. S. and Saslaw, W. C.},
  title   = {21-cm absorption in NGC 1275},
  journal = {The Astrophysical Journal},
  year    = {1973},
  volume  = {185},
  pages   = {809--815},
  doi     = {10.1086/152456},
  eprint = {https://ui.adsabs.harvard.edu/abs/1973ApJ...185..809D/abstract}
}

@article{Ekers1976,
  author  = {Ekers, R. D. and van der Hulst, J. M. and Miley, G. K.},
  title   = {The 21-cm absorption line in NGC1275},
  journal = {Nature},
  year    = {1976},
  volume  = {262},
  pages   = {369--370},
  doi     = {10.1038/262369a0}
}

@ARTICLE{Gupta2017,
       author = {Gupta, N. and Srianand, R. and Baan, W. and {et al.}},
        title = "{The MeerKAT Absorption Line Survey (MALS): Survey Design and First Results}",
      journal = {Proceedings of Science, Workshop on "MeerKAT Science: On the Pathway to the SKA"},
         year = 2017,
       volume = {},
        number = {},
        pages = {},
          doi = {10.48550/arXiv.1708.07371},
       archivePrefix = {arXiv},
       eprint = {https://doi.org/10.48550/arXiv.1708.07371}
}

@ARTICLE{Curran2024,
       author = {{Curran}, S. J.},
        title = "{The depletion of star-forming gas by AGN activity in radio sources}",
      journal = {Publications of the Astronomical Society of Australia},
         year = 2024,
        month = feb,
       volume = {41},
        number = {},
          doi = {10.1017/pasa.2024.1},
       eprint = {arXiv:2401.00962},
       archivePrefix = {arXiv},
       adsurl = {https://doi.org/10.1017/pasa.2024.1}
}

@article{Aditya2018,
  author  = {Aditya, J. N. H. S. and Kanekar, Nissim},
  title   = {A Giant Metrewave Radio Telescope survey for associated H\,{\sc i} 21\,cm absorption in the Caltech--Jodrell flat-spectrum sample},
  journal = {Monthly Notices of the Royal Astronomical Society},
  year    = {2018},
  volume  = {481},
  number  = {2},
  pages   = {1578--1596},
  doi     = {https://doi.org/10.1093/mnras/sty2184}
}

@ARTICLE{Aditya2024,
       author = {Aditya, J. N. H. S. and Yoon, Hyein and Allison, James R. and others},
        title = "{The FLASH pilot survey: an H I absorption search against MRC 1-Jy radio sources}",
      journal = {Monthly Notices of the Royal Astronomical Society},
         year = 2024,
        month = jan,
       volume = {527},
        number = {3},
        pages = {8511--8534},
          doi = {10.1093/mnras/stad3722},
       eprint = {arXiv:2310.14571},
       archivePrefix = {arXiv},
       adsurl = {https://doi.org/10.1093/mnras/stad3722}
}

@ARTICLE{Li2024,
       author = {{Li}, Yingjie and {Xu}, Ye and {Li}, Jingjing and {Bian}, Shuaibo and {Lin}, Zehao and {Hao}, Chaojie and {Liu}, Dejian},
        title = "{VLBI with SKA: Possible Arrays and Astrometric Science}",
      journal = {Research in Astronomy and Astrophysics},
         year = 2024,
        month = apr,
       volume = {},
        number = {},
          doi = {https://arxiv.org/pdf/2404.14663},
       adsurl = {https://arxiv.org/pdf/2404.14663}
}

@ARTICLE{krumholz2013,
       author = {{Krumholz}, M.~R.},
        title = "{The big problems in star formation: The star formation rate, stellar clustering, and the initial mass function}",
      journal = {Physics Reports},
         year = 2014,
        month = Jun,
       volume = {539},
        pages = {49-134},
          doi = {10.48550/arXiv.1402.0867},
       adsurl = {https://arxiv.org/abs/1402.0867}
}

@ARTICLE{leroy2008,
       author = {{Leroy}, A.~K. and {Walter}, F. and {Brinks}, E. and et al.},
        title = "{The Star Formation Efficiency in Nearby Galaxies: Measuring Where Gas Forms Stars Effectively}",
      journal = {The Astronomical Journal},
         year = 2008,
        month = dec,
       volume = {136},
        pages = {2782-2845},
          doi = {10.1088/0004-6256/136/6/2782},
       adsurl = {https://ui.adsabs.harvard.edu/abs/2008AJ....136.2782L}
 }

@article{Silk1998,
  author    = {Silk, Joseph and Rees, Martin J.},
  title     = {Quasars and galaxy formation},
  journal   = {Astronomy and Astrophysics},
  volume    = {331},
  pages     = {L1--L4},
  year      = {1998},
  eprint = {astro-ph/9801013},
  archivePrefix = {arXiv},
  primaryClass = {astro-ph},
  adsurl       = {https://ui.adsabs.harvard.edu/abs/1998A\&A...331L...1S/abstract}
}

@article{Fabian2012,
  author    = {Fabian, Andrew C.},
  title     = {Observational Evidence of Active Galactic Nuclei Feedback},
  journal   = {Annual Review of Astronomy and Astrophysics},
  volume    = {50},
  number    = {1},
  pages     = {455--489},
  year      = {2012},
  doi       = {10.1146/annurev-astro-081811-125521},
  archivePrefix= {arXiv},
  adsurl       = {https://ui.adsabs.harvard.edu/abs/2012ARA%26A..50..455F/abstract}
}

@article{Heckman2014,
  author    = {Heckman, Timothy M. and Best, Philip N.},
  title     = {The Coevolution of Galaxies and Supermassive Black Holes: Insights from Surveys of the Contemporary Universe},
  journal   = {Annual Review of Astronomy and Astrophysics},
  volume    = {52},
  pages     = {589-660},
  year      = {2014},
  doi       = {10.1146/annurev-astro-081913-035722},
  archivePrefix= {arXiv},
  adsurl       = {https://ui.adsabs.harvard.edu/abs/2014ARA%26A..52..589H/abstract}
}

@article{Morganti2004,
  author       = {Morganti, R. and Oosterloo, T. A. and Tadhunter, C. N. and Vermeulen, R. and Pihlström, Y. M. and van Moorsel, G. and Wills, K. A.},
  title        = {The unfriendly ISM in the radio galaxy 4C 12.50 (PKS 1345+12)},
  journal      = {Astronomy \& Astrophysics},
  year         = {2004},
  volume       = {424},
  pages        = {119--124},
  doi          = {10.1051/0004-6361:20041064},
  publisher    = {EDP Sciences},
  url          = {https://doi.org/10.1051/0004-6361:20041064}
}

@article{Wagh2024,
  author       = {Wagh, S. and Pandey-Pommier, M and Roy, Nirupam and others},
  title        = {Exploring neutral hydrogen in the radio Molecular Hydrogen Emission Galaxies (MOHEGs) and prospects with the SKA},
  journal      = {The Astrophysical Journal},
  year         = {2024},
        month = Mar,
       volume = {963},
       number = {2},
        pages = {11pp},
          doi = {10.3847/1538-4357/ad1edf},
archivePrefix = {arXiv},
  eprint       = {2401.07613},
  archivePrefix= {arXiv},
  primaryClass = {astro-ph.GA},
  adsurl       = {https://arxiv.org/abs/2401.07613}
}

@article{Ogle2010,
  author = {Ogle, P. M. and Boulanger, F. and Guillard, P. and Evans, D. A. and Antonucci, R. and Appleton, P. N. and Nesvadba, N. and Leipski, C.},
  title        = {Jet-powered molecular hydrogen emission from radio galaxies},
  journal      = {Astrophysical Journal},
  year         = {2010},
        month = Nov,
       volume = {724},
       number = {2},
        pages = {1193-1209},
          doi = {10.1088/0004-637X/724/2/1193},
  archivePrefix= {arXiv},
  primaryClass = {astro-ph.GA},
  adsurl       = {https://arxiv.org/abs/1009.4533}
}

@article{Morganti2023,
  author       = {Morganti, Raffaella and Murthy, Suma and Oosterloo, Tom and Blanchard, Jay and Cook, Claire and Paragi, Zsolt and Orienti, Monica and Nagai, Hiroshi and Schulz, Robert},
  title        = {Cold gas in the heart of Perseus A},
  journal      = {Astronomy \& Astrophysics},
  year         = {2023},
  volume       = {678},
  pages        = {A42},
  doi          = {10.1051/0004-6361/202347117},
  url          = {https://doi.org/10.1051/0004-6361/202347117}
}

@article{Oosterloo2023,
  author       = {Oosterloo, Tom and Morganti, Raffaella and Murthy, Suma},
  title        = {Closing the feedback–feeding loop of the radio galaxy 3C 84},
  journal      = {Nature Astronomy},
  year         = {2023},
  volume       = {7},
  number       = {12},
  pages        = {1506--1513},
  doi          = {10.1038/s41550-023-02138-y},
  url          = {https://doi.org/10.1038/s41550-023-02138-y}
}

@article{Schulz2018,
  author = {Schulz, R. and Morganti, R. and Nyland, K. and Oosterloo, T. and Harwood, J. and Mukherjee, D.},
  title = {Cold gas and radio jets: VLBI imaging of HI absorption in the restarted radio galaxy 3C 236},
  journal = {Monthly Notices of the Royal Astronomical Society},
  year = {2018},
  volume = {617},
  number = {A38},
  doi = {https://doi.org/10.1051/0004-6361/201833108 }
}

@article{Vermeulen2003,
  author = {Vermeulen, R. C. and Pihlström, Y. M. and Tschager, W. and de Vries, W. H. and others},
  title = {Observations of H I absorbing gas in compact radio sources
at cosmological redshifts},
  journal = {Astronomy \& Astrophysics},
  year = {2003},
  volume = {404},
  pages = {861--875},
  doi = {10.1051/0004-6361:20030468}
}

@ARTICLE{Maccagni2017,
       author = {{Maccagni}, F.M. and {Morganti}, R. and {Oosterloo}, T.~A. and {Ger{\'e}b}, K. and {Maddox}, N.},
        title = "{Kinematics and physical conditions of H I in nearby radio sources. The last survey of the old Westerbork Synthesis Radio Telescope}",
      journal = {A\&A},
         year = 2017,
        month = aug,
       volume = {604},
          eid = {A43},
        pages = {A43},
          doi = {10.1051/0004-6361/201730563},
archivePrefix = {arXiv},
       eprint = {1705.00492},
 primaryClass = {astro-ph.GA},
       adsurl = {https://ui.adsabs.harvard.edu/abs/2017A\&A...604A..43M}
}

@ARTICLE{Maccagni2023,
       author = {Maccagni, F. M. and Ruffa, I. and Loni, A. and Prandoni, I. and others},
        title = "{The AGN fuelling/feedback cycle in nearby radio galaxies - V. The cold atomic gas of NGC 3100 and its group}",
      journal = {Astronomy \& Astrophysics},
         year = 2023,
       volume = {675},
        pages = {A59},
          doi = {10.1051/0004-6361/202346521}
}

@article{Aditya2019,
  author = {Aditya, J. N. H. S.},
  title = {uGMRT detections of HI 21-cm absorption associated with intermediate redshift galaxies},
  journal = {Monthly Notices of the Royal Astronomical Society},
  year = {2019},
  volume = {482},
  number = {4},
  pages = {5597--5612},
  doi = {https://academic.oup.com/mnras/article/482/4/5597/5173102}
}

@article{Morganti2005,
  author = {Morganti, R. and Tadhunter, C. N. and Oosterloo, T. A.},
  title = {Fast neutral outflows in powerful radio galaxies: a major source of feedback in massive galaxies},
  journal = {Astronomy \& Astrophysics},
  year = {2005},
  volume = {444},
  number = {1},
  pages = {L9--L13},
  doi = {10.1051/0004-6361:200500197}
}

@article{Morganti2016,
  author = {Morganti, R. and Veilleux, Sylvain and others},
  title = {Another piece of the puzzle: The fast H I outflow in Mrk 231 },
  journal = {Astronomy \& Astrophysics},
  year = {2016},
  volume = {593},
  pages = {11},
  doi = {10.1051/0004-6361/201628978}
}

@article{Morganti2003,
  author = {Morganti, R. and Oosterloo, T. A.  and others},
  title = {Fast Outflow of Neutral Hydrogen in the Radio Galaxy 3C 293 },
  journal = {The Astrophysical Journal},
  year = {2003},
  volume = {593},
  pages = {L69-L72},
  doi = {10.1086/378219}
}

@article{Oosterloo2025,
  author = {Oosterloo, T. A. and Morganti, R. and Tadhunter, C. and others},
  title = {The changing impact of radio jets as they evolve: The view from the cold gas},
  journal = {A\&A},
  year = {2025},
  volume = {700},
  number = {1},
  doi = {10.1051/0004-6361/202554536 }
}

@article{Wiklind1995,
  author    = {Wiklind, Tommy and Henkel, Christian},
  title     = {Cold dust in elliptical galaxies},
  journal   = {Astronomy and Astrophysics},
  volume    = {297},
  pages     = {L71-L74},
  year      = {1995},
  doi       = {https://ui.adsabs.harvard.edu/scan/manifest/1995A\&A...297L..71W}
}

@article{Cicone2014,
  author    = {Cicone, Claudia and Maiolino, Roberto and Sturm, Emanuele and Graciá-Carpio, Javier and Feruglio, Chiara and Neri, Roberto and Aalto, Susanne and Davies, Ric and García-Burillo, Santiago and González-Alfonso, Eduardo and Hailey-Dunsheath, Steven and Piconcelli, Enrico and Veilleux, Sylvain},
  title     = {Massive Molecular Outflows and Evidence for AGN Feedback from CO Observations},
  journal   = {Astronomy \& Astrophysics},
  volume    = {562},
  pages     = {A21},
  year      = {2014},
  doi       = {10.1051/0004-6361/201322464}
}

@article{Laing2014,
  author    = {Laing, R. A. and Bridle, A. H.},
  title     = {Systematic Properties of Decelerating Relativistic Jets in Low-luminosity Radio Galaxies},
  journal   = {Monthly Notices of the Royal Astronomical Society},
  volume    = {437},
  number    = {4},
  pages     = {3405--3441},
  year      = {2014},
  doi       = {10.1093/mnras/stt2138}
}

@article{Veilleux2005,
  author    = {Veilleux, Sylvain and Cecil, Gerald and Bland-Hawthorn, Jonathan},
  title     = {Galactic Winds},
  journal   = {Annual Review of Astronomy and Astrophysics},
  volume    = {43},
  pages     = {769--826},
  year      = {2005},
  doi       = {10.1146/annurev.astro.43.072103.150610}
}

@article{Tadhunter2016,
  author    = {Tadhunter, Clive},
  title     = {Radio AGN in the local universe: unification, triggering and evolution},
  journal   = {Astronomy and Astrophysics Review},
  volume    = {24},
  pages     = {10},
  year      = {2016},
  doi       = {10.1007/s00159-016-0094-x}
}

@article{Morganti2017,
  author    = {Morganti, R.},
  title     = {The Many Routes to AGN Feedback},
  journal   = {Frontiers in Astronomy and Space Sciences},
  volume    = {4},
  pages     = {42},
  year      = {2017},
  doi       = {10.3389fspas.2017.00042}
}

\end{document}